\documentclass[journal=jacsat,manuscript=article]{achemso}

\usepackage[version=3]{mhchem} 
\usepackage{graphicx}
\usepackage{epstopdf}

\author{Claudio Melis}
\affiliation[University of Cagliari]{ Dipartimento di Fisica Universit\`a di Cagliari}
\alsoaffiliation[SLACS CNR-INFM]{ Sardinian Laboratory for Computational Materials Science, SLACS (CNR-INFM), Cittadella Universitaria, I-09042 Monserrato (Ca), Italy}
\email{claudio.melis@dsf.unica.it}

\author{Alessandro Mattoni}
\affiliation[SLACS CNR-INFM]{ Sardinian Laboratory for Computational Materials Science, SLACS (CNR-INFM), Cittadella Universitaria, I-09042 Monserrato (Ca), Italy}

\author{Luciano Colombo}
\affiliation[University of Cagliari]{ Dipartimento di Fisica Universit\`a di Cagliari}
\alsoaffiliation[SLACS CNR-INFM]{ Sardinian Laboratory for Computational Materials Science, SLACS (CNR-INFM), Cittadella Universitaria, I-09042 Monserrato (Ca), Italy}

\title{Atomistic investigation of the poly(3-hexylthiophene) adhesion on nanostructured titania}

\begin{document}
\begin{abstract}
 We study the adhesion of poly(3-hexylthiophene) on nanostructured titania surface in vacuo by means of model potential 
 molecular dynamics. 
 We generate large scale atomistic models of nanostructured titania surfaces (consisting of spherical nanocaps on top of a (110) rutile surface)  and
 we study the adhesion of an oligothiophene as a function of their local curvature and
 roughness.
 In the limit of a perfect planar surface, the maximum adhesion energy is calculated to be as large as $0.6$ eV per monomer, 
 and it corresponds to the oligothiophene oriented along the [$\bar{1}10$] direction of the surface. Deformations of the polymer 
 are observed due the incommensurability between the titania and the polymer lattice parameters.
 When the surface is nanostructured, the adhesion of the polymer is affected by the local morphology and a nonmonotonic dependence on the
 surface curvature is observed. 
 The atomistic results are explained by a simple 
 um model that includes the strain energy of the polymer and
 its electrostatic interaction with the local surface charge.
 \end{abstract}

\section{Introduction}
Polymer based hybrid nanomaterials formed by a polymer interfaced to an inorganic substrate have large impact in
modern materials science for  applications either
as structural materials\cite{struct-pol} (matrices for high performance composites\cite{struct-pol1}) 
or as functional materials\cite{struct-pol2} ( catalyst supports\cite{struct-pol3} and microelectronic devices\cite{struct-pol4}).
Among many others, a critical issue of hybrid systems concerns
the adhesion of its 
organic and inorganic components, 
 critically affecting  the resulting  mechanical, thermal and optoelectronic properties. 
The adhesion strength, in turn,  
depends mostly  on the 
the interface chemistry and the atomic scale morphology.
 
 As a matter of fact, the atomic scale understanding of the adhesion is a matter of debate.
Adhesion can be  the result of several interatomic force actions including  covalent, electrostatic and dispersive (Van der Waals)
ones\cite{pol-adhes}.
The relevance of each contribution depends both on the chemistry and on the atomic scale structural properties. 
It is not clear, for example, whether an intense macroscopic adhesion 
is always the result of strong covalent bonding at the interface.
More in details: when considering a polymer/metaloxyde hybrid, the covalent bonds are not expected to be the major contribution to adhesion since, in general,
the polymer does not form covalent bonds with the inorganic material.
Nevertheless, intense electrostatic interactions occur between the ions of the surface and the partially charged atoms
 in the polymers due to the ionicity of the metaloxide.
 This is, for instance, the case of  poly3-hexylthiophene (poly3HT) for which large atomic partial charges 
 (up to $0.15e$, where $e$ is the electronic charge) are found.
 Accordingly, a comparatively strong adhesion between P3HT and titania is expected, due to contributions other than covalent binding. 
 
A second important issue to take into account is
  the effect of the morphology of the nanostructured substrate films like, e.g., titania films formed by
cluster assembling \cite{titania}
or zincoxyde nanostructures \cite{ackermann}. 
Because of their large surface-to-volume ratio, such nanostructured films are expected to improve the polymer adhesion giving rise to heterojunctions with large interface area.
 To date it is not clear  how the curvature at the nanoscale can affect the adhesion strength.

Besides the inorganic substrate, it is also
 crucial to consider the actual morphology of the  polymer.
A polymer chain can be largely distorted as a result of its interaction with   
the surface, if its periodicity is incommensurable with the surface lattice structure.
Accordingly, it is important to take into account the large strain energy associated to the distortions to properly model the adhesion phenomena.
A reasonable expectation is that the adhesion is the result of  the balance between the formation of the largest number
of favourable electrostatic interactions, and the minimization of the strain energy associated to distortions of the polymer.

The  polymer/substrate adhesion also affects the overall efficiency of polymer based solar cells. 
Polymer based hybrids  (e.g. P3HT/TiO$_2$,  P3HT/ZnO \cite{gunes}) have emerged as  promising systems for photovoltaics, since they
 can in principle combine 
 the good formability of polymers and the good trasport properties and thermal stability of the inorganic metaloxide.
A strong link of the polymer (where light is absorbed) to the inorganic substrate (where electrons
are accepted) is necessary  
 to give rise to an efficient photoconversion.
Accordingly, the theoretical understanding of the interface structure at the atomic scale (i.e. the actual interatomic distances, the overlap of the electronic density between polymer and substrate, the covalent versus electrostatic nature of the bonding)
  is of great relevance to improve the properties of such hybrid materials. 

The above scenario underlines the present work, which is focused on poly-(3hexylthiophene)/TiO$_2$  here selected as a prototype of hybrid interface.
Despite the technological relevance of the system \cite{P3HT-TIO2}, there is a poor knowledge 
about the bonding chemistry and structural properties of its
 interface. We generate large scale atomistic models of the P3HT /TiO$_2$ interface and we study the polymer adhesion  under ideal conditions of 
 chemical and structural purity.
 We focus on the rutile phase of titania that is the most common form in nature\cite{greenwood}. 
In particular, we study how the curvature at the nanoscale affects the local  titania-molecule adhesion strength.
In order to represent  the structural complexity associated to nanostructured films and 
polymer distortions, we make use of 
models including up to  $10^4$ atoms.
By using molecular dynamics (MD) simulations we are able to extensively explore the attraction basin between the polymer and the titania, as well as to calculate the
adhesion energy as a function of its 
curvature and roughness.
Since the system size here investigated falls out of reach of a systematic first principle calculation, the interatomic forces are derived from model potentials (MP).

\section{Theoretical framework}

The description of interatomic forces in hybrids is challenging.
A general model potential for the hybrid system  is not available, while there exist reliable potentials
for titania or polymers, separately. 
Here we combine such existing force fields by adding long range Coulomb and dispersive interactions 
to model interactions across the metaloxyde-polymer boundary.
The model is  validated against experiments and first-principles calculations. 
 
The  TiO$_2$ rutile was described by the sum of a Coulomb and a Buckingham-type two-body potential\cite{buckingham} of the form : 
\begin{equation} \label{eqn:buckingham}
U(r_{ij})= \frac{q_iq_j}{r_{ij}}+Ae^{(-r_{ij}/\rho)} - \frac{C}{r_{ij}^6}
\end{equation}
which has been extensively used for metal oxides\cite{buckingham1}. 
Here q$_i$ and q$_j$ are the charges of atoms $i$ and $j$ while $r_{i j}$ is their relative distance. The first term in equation \ref{eqn:buckingham} takes into account  long-range Coulomb interactions, the second term is a short range repulsion potential, the third term is the van der Waals attraction.
The parameters   A, $\rho$ and C, and all the atomic charges  q$_i$ were taken from reference\cite{buckingham}. 
The  lattice parameters  {\itshape a} and {\itshape c}
of the rutile \cite{exp-rutile} crystal structure calculated according to the present model are
in fairly agreement with the experiments (errors are within 1.7\% and  2.3\% for  {\itshape a} and {\itshape c} , respectively). 

In order to describe P3HT we used the AMBER force field\cite{AMBER}, that includes either bonding (bonds, bending, torsional) and non-bonding (van der Waals plus Coulomb) contributions.
The atomic partial charges were calculated according to the standard  AM1-CM2 method\cite{am1-cm2}. 
The  AMBER force field  was validated against  experiments\cite{THIO-EXP} on a  single thiophene molecule for which experimental data are available. 
The geometry of thiophene  was fully relaxed  at  MP level  using the conjugate gradients method.
It is found that the structural parameters (bond lengths and angles) of the molecule are in good agreement with the experiments (deviations are less than $4$\%) confirming the 
reliability of this force field for the description of oligothiophenes.

Finally, for the  P3HT/TiO$_2$ interaction we used an interatomic potential consisting in  the sum of all $i-j$ atomic pairs 
( $i$ and $j$ running, respectively, over all the polymer atoms and the titania film atoms) of the form:
\begin{equation}
 U(r_{ij})= \frac{q_iq_j}{r_{ij}}+4\epsilon_{ij}\left[\left(\frac{\sigma_{ij}}{r_{ij}}\right)^{12} - \left(\frac{\sigma_{ij}}{r_{ij}}\right)^{6}\right]
\end{equation}
The first term is the Coulomb contribution due to the atomic partial charges
(which are the same as above).
 The Lennard-Jones parameters $\sigma_{ij}$ and $\epsilon_{ij}$ for the mixed TiO$_2$-P3HT interaction were obtained  
   from the values ($\sigma_{ii}, \epsilon_{ii}$) for like-atoms pairs and setting $\epsilon_{ij}=\sqrt{\epsilon_{ii} \epsilon_{jj} }$ and  $\sigma_{ij} =\frac{\sigma_{ii}+\sigma_{jj}}{2}$ for unlike-atoms ones. The parameters corresponding to ($i$-$i$) were taken from  Ref.\cite{sushko} in the case of  Ti and from the AMBER database for Oxygen, Carbon, Hydrogen and Sulfur\cite{AMBER}.

The reliability of the above model potential for P3HT/TiO$_2$ was 
validated by studying the interaction of a single thiophene molecule on a TiO$_2$ surface. This is a stringent test case since a
small thiophene molecule is the building block of P3HT and its aromatic $\pi$ ring is responsible for the optical activity of the polymer. 
First principles calculations\cite{THIO-DFT}(based on the CASTEP energy package\cite{CASTEP,CASTEP1}) have been performed for the
adsorption of a thiophene on a rutile $(110)$ surface.
The above model potential is able to quantitatively reproduce the first principle results for both
the geometry and the adhesion energy of the thiophene (see next Section).
Such an agreement is related to the non covalent nature of the binding between aromatic molecules and
rutile $(110)$ surface, as discussed in the literature\cite{ACENES-TIO2}.

 All the simulations were performed by using DL\_POLY\cite{dlpoly}(version 2.19). 
Atomic trajectories were calculated by the velocity Verlet algorithm, with a time step of 1 fs.
Long-range Coulomb interactions were evaluated using a  particle mesh Ewald algorithm\cite{ewald}.
The convergence parameter was set to $4.0$ \AA$^{-1}$ with a mesh of $31$ x $48$ x $19$ wavevectors in the x, y, and z directions, respectively. Van der Waals interactions were cutoff at $9.5$ \AA \ . As for the interatomic model potential, we combined existing model potentials as detailed below.  
The simulation cells contained up to 12202 atoms in total.

\section{Results and discussion}\label{RES}

\subsection{Equilibrium properties of P3HT}
\begin{figure}
\includegraphics[scale=1.0]{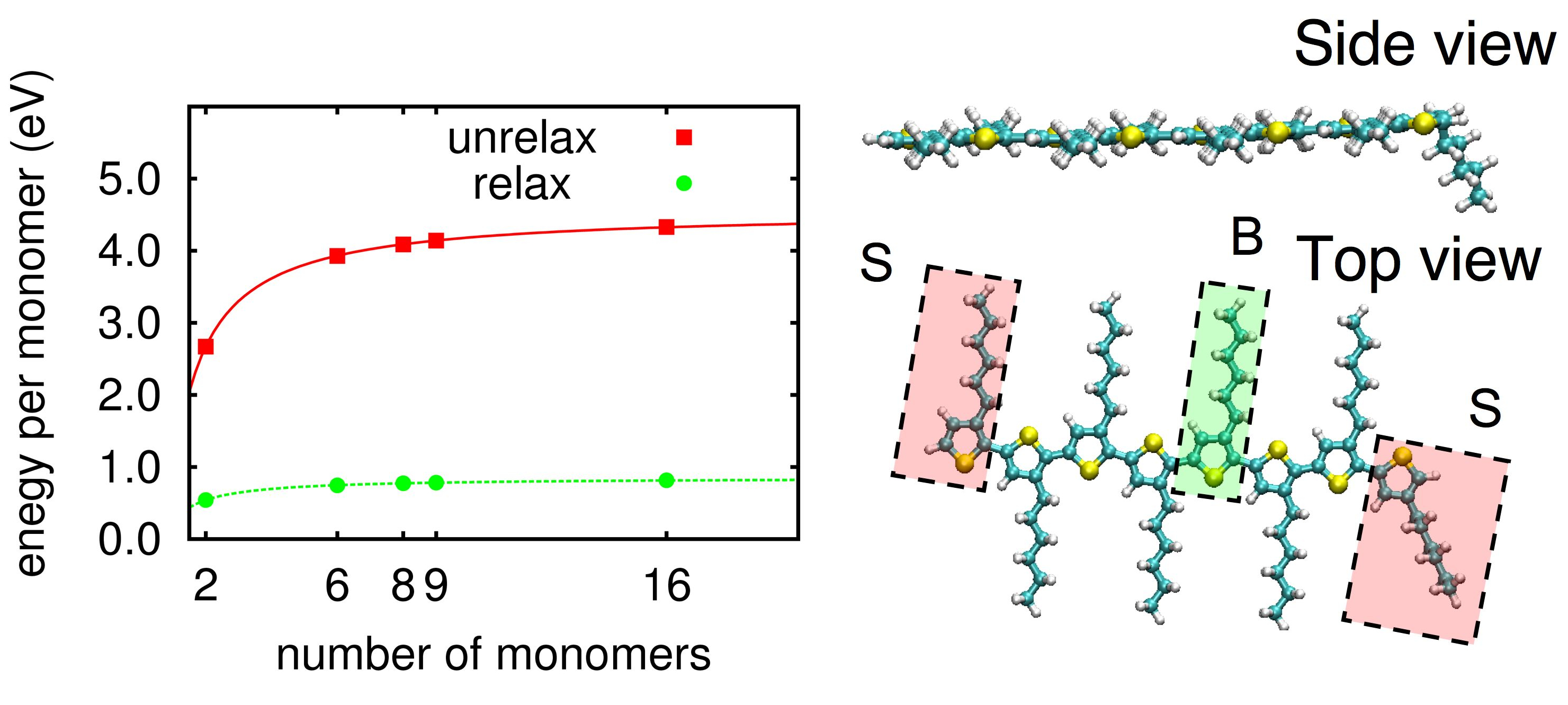}
\caption{  \label{poly_curve}
(Right) Stick and balls representation of the equilibrium structure (in vacuo) of 3-hexylthiophene chain (eight monomers).
The energy and the geometry of inner monomers (B) differ from monomers at the boundaries (S).
Carbon, Hydrogen and  Sulphur atoms are represented in 
cyan, white and yellow, respectively. 
(Left) Energy (normalized to the number of monomers) of a 3-hexylthiophene chain  as a function of the its length.
Full squares and full circles correspond to the energy of the unrelaxed and relaxed structures, respectively. Lines 
correspond to the model $\epsilon(N)$, see text.}
\end{figure}

We investigated firstly the equilibrium configuration of an oligothiophene (N-3hexylthiophene, N-3HT) formed by 
a sequence of $N$  units of 3HT. 
The polymer chain corresponds to the case of large number of units $N$. 
The geometries were locally relaxed in vacuo 
(see \ref{poly_curve}, right)
by starting from  planar configurations, where all the 3HT monomers lies in the same plane. 
We generated oligomers corresponding to lengths $2 \le N \le 16$.
  The dangling bonds at the boundaries were  saturated by hydrogen atoms.
We point out  that  in order to saturate the boundaries, other groups are possible (e.g. methyl),
 but for the purposes of the present analysis the actual choice is unimportant.
 In fact,  in the limit of long chains the interaction is dominated by the chemistry of the inner monomers.
 Furthermore, the effect of the substrate morphology on the P3HT/substrate binding is calculated as differences
 between structures with the same saturation group.
 
During a local minimization, based on conjugated gradient algorithm, the planar geometry is
preserved in any case, regardless the actual chain length ($N$). The corresponding structure 
is referred to as the unrelaxed chain. 
The energy per monomer $\epsilon(N)$ as a function of $N$ is reported in \ref{poly_curve} (left) as full squares.
The unrelaxed models were further optimized through a two-step minimization precedure:
(i) $10$-ps long low temperarature annealing ($10$ K);
(ii) another geometry optimization with a conjugate-gradient algorithm.
Asymmetric deviations from the ideal planar geometry are observed in the final geometries (referred to as relaxed), mostly involving the hexyl chains  and due to steric interaction.
An example of N-3HT is reported in \ref{poly_curve} (right) for the case $N=8$.  
Upon the relaxation procedure, a large energy decrease is observed ($\sim$3 eV)  for all the chains (full circles in left panel of
\ref{poly_curve}).

Both the unrelaxed and relaxed data exhibit a dependence on the chain length.
In particular it is found that the energy per monomer increases monotonically with $N$.
A simple model can be obtained as follows:
let $\epsilon_S$ and $\epsilon_B$ the energy corresponding to the S and B monomers (see \ref{poly_curve}) in N-3HT chain.
$\epsilon_S$ is obtained from $\epsilon_B$ by adding the energy necessary to remove half a $C-C$ bond (each monomer
shares a C-C bond with its neighbor) and subtracting the energy gained  by forming a $C-H$ bond.
Accordingly, we can write $\epsilon_S=\epsilon_B - \Delta$ where
$\Delta = |E_{C-H}|-0.5|E_{C-C}|$ is a positive number.
The total energy  $E(N)$ of an N-3HT chain is calculated as 
\begin{equation}
E(N)= 2\epsilon_S + (N-2)\epsilon_B
\end{equation}
The energy per monomer $\epsilon(N)=E(N)/N$ is a function of the length $N$
\begin{equation}
\label{energy }
\epsilon(N)=\epsilon_B - \frac{2\Delta}{N}
\end{equation}
where $\Delta$ and $\epsilon_B$ can be obtained by fitting the function $\epsilon(N)$ on the atomistic data.
In the limit of an infinite chain ($N$ $\rightarrow$ $\infty$) we get  $\epsilon (\infty)= \epsilon_B$.
In the opposite limit $N=2$, it is correctly found $\epsilon(2)=\epsilon_B - \Delta=\epsilon_S$.
This model reproduces accurately the atomistic data for both the unrelaxed and  relaxed structures  
(red and green lines of \ref{poly_curve}, respectively).    
The corresponding  quantities
 will be hereafter labeled by the letter $u$ and $r$, respectively.
Present calculations provided  $\Delta^u \sim 1.9 $ eV and 
$ \Delta^r \sim 0.31$ eV.

We further observe (see left panel of \ref{poly_curve} ) that $\epsilon^u(\infty)-\epsilon^r(\infty) >\epsilon^u(2) - \epsilon^r(2)$.
Since  
 $\epsilon^u (\infty)-\epsilon^r (\infty) = \epsilon_B^u - \epsilon_B^r $ and 
$\epsilon^u(2) - \epsilon^r(2)= \epsilon^u_S - \epsilon^r_S$, we conclude that $ \epsilon^u_S - \epsilon^r_S <   \epsilon^u _B-\epsilon^r _B $, i.e. during the molecule relaxation the  energy decrease  at the boundaries (S) is the smallest.

\subsection{P3HT/TiO$_2$ adhesion}
As a first step towards the study of P3HT/TiO$_2$ interaction we focus on the case of a thiophene molecule on a $(110)$ rutile surface.
Such a system is of technological interest per se (for titania catalyzed desulfurization  of thiopene\cite{DESULF}) and it is, at the
same time, representative of the polymer.

We start by generating a model for a planar titania surface. The surface with the lowest formation energy
  is obtained by cutting along a $(110)$ plane in such a way that only the Ti-O bonds along the $[110]$
direction are cut. 
The $(110)$ surface consists of two-coordinated oxygen ions (forming the external rows along the $[001]$ directions) bonded to the six- coordinated titanium ions which alternate with five- coordinated titanium ions along the $[\bar{1}10]$.
The relaxed (unrelaxed) surface energy is calculated to be $0.110$ eV\AA$^{-2}$ ( $0.137$ eV\AA$^{-2}$ ).
Different $(110)$ cuts give rise to unfavourable surfaces with a large density of dangling bonds and, in turn,  to a much larger surface energy.

\begin{figure}
\includegraphics[scale=0.4]{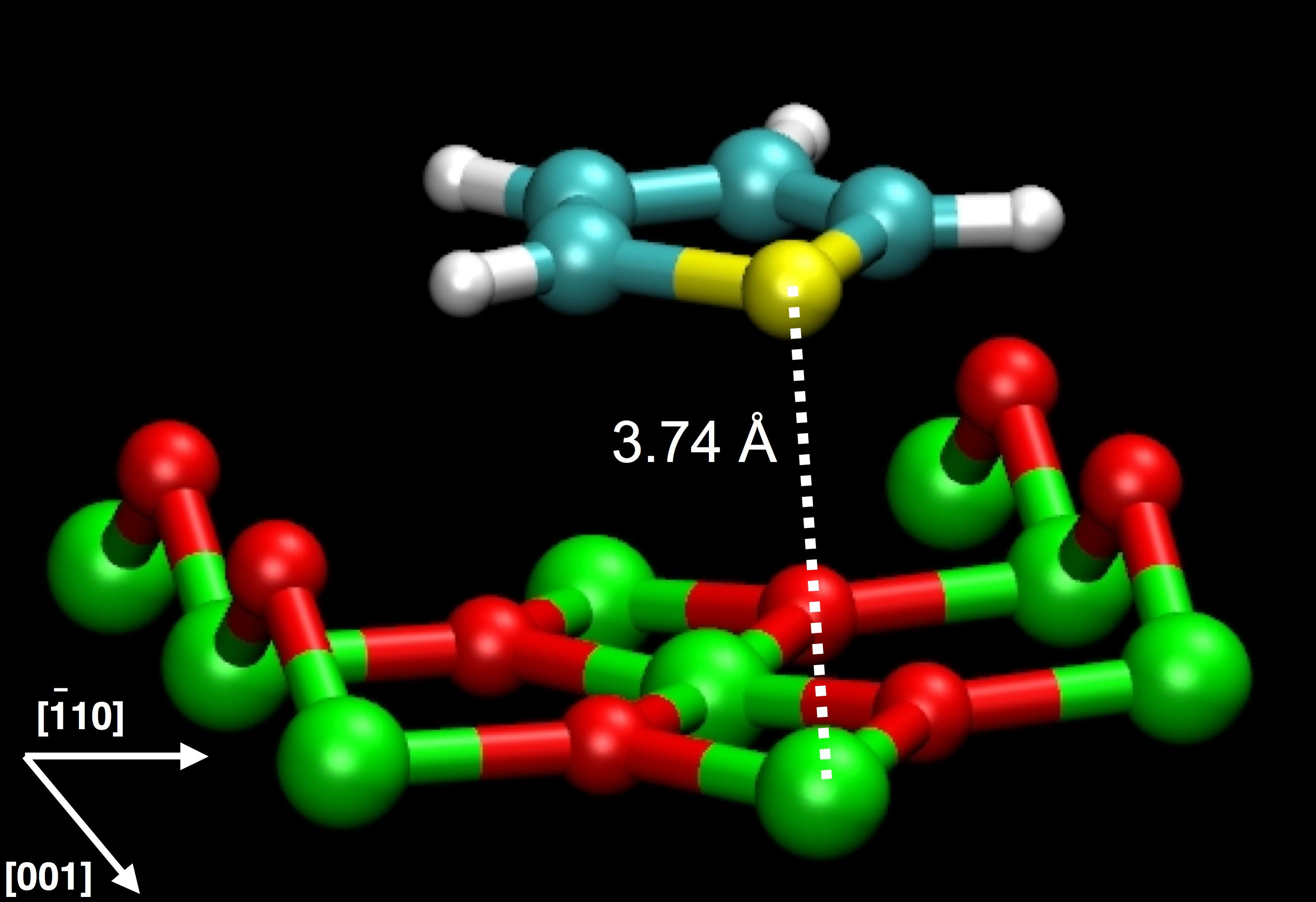}
\caption{  \label{THIO-TIO2}
 Stick and balls representation of the thiophene/TiO$_2$ rutile (110) bound state. Carbon, Hydrogen, Sulphur, Oxygen and Titanium atoms are represented in 
cyan, white, yellow, red and green, respectively. 
}
\end{figure}

In order to study the interaction with the thiophene, 
we  placed the molecule  on top of  the surface   at a distance of $\sim 1$ nm
(much larger than corresponding values in the  bound configuration), and we performed low temperature MD run as 
long as $100$-ps, then followed by a geometry optimization with the conjugate-gradient algorithm.
As a result of the long range Coulomb and dispersion forces, the aromatic molecule is attracted by the surface untill the 
bound state is formed.
The  minimum energy bound configuration is represented in \ref{THIO-TIO2}, where the thiophene ring is parallel
to the surface and lies between to oxygen rows with its C-S-C tip oriented along the $[001]$ crystallographic direction.
The S atom (yellow)  and the center of the thiophene are on top of two adjacent undercoordinated Ti atoms.
This configuration is in good agreement with the predictions by DFT results\cite{THIO-DFT}.
 The MP calculated S-Ti  distance ($3.74$ \AA ) is within 10 \% from the DFT value, while the C-S bonds are
only $2$ \% longer. 
Notably, MP and DFT calculations  give the same adsorption energy equal to $0.52$ eV.
The accuracy of the MP results  is mainly due to  the non-covalent nature of the thiophene-rutile binding.
This is demostrated by the large molecule-surface distance which prevents a sizable overlap between the
 the $d$ orbitals of titanium atoms and the $\pi$ electrons of the thiophene ring.
The binding is rather dominated by the Coulomb contribution (e.g. the 
oxygen-carbon electrostatic repulsions) that are correctly described by
 the model potential.
Similar conclusions are valid also for a large class of aromatic molecules (e.g. acenes) on TiO$_2$ $(110)$ rutile surface\cite{ACENES-TIO2}.

\begin{figure}
\includegraphics[scale=0.2]{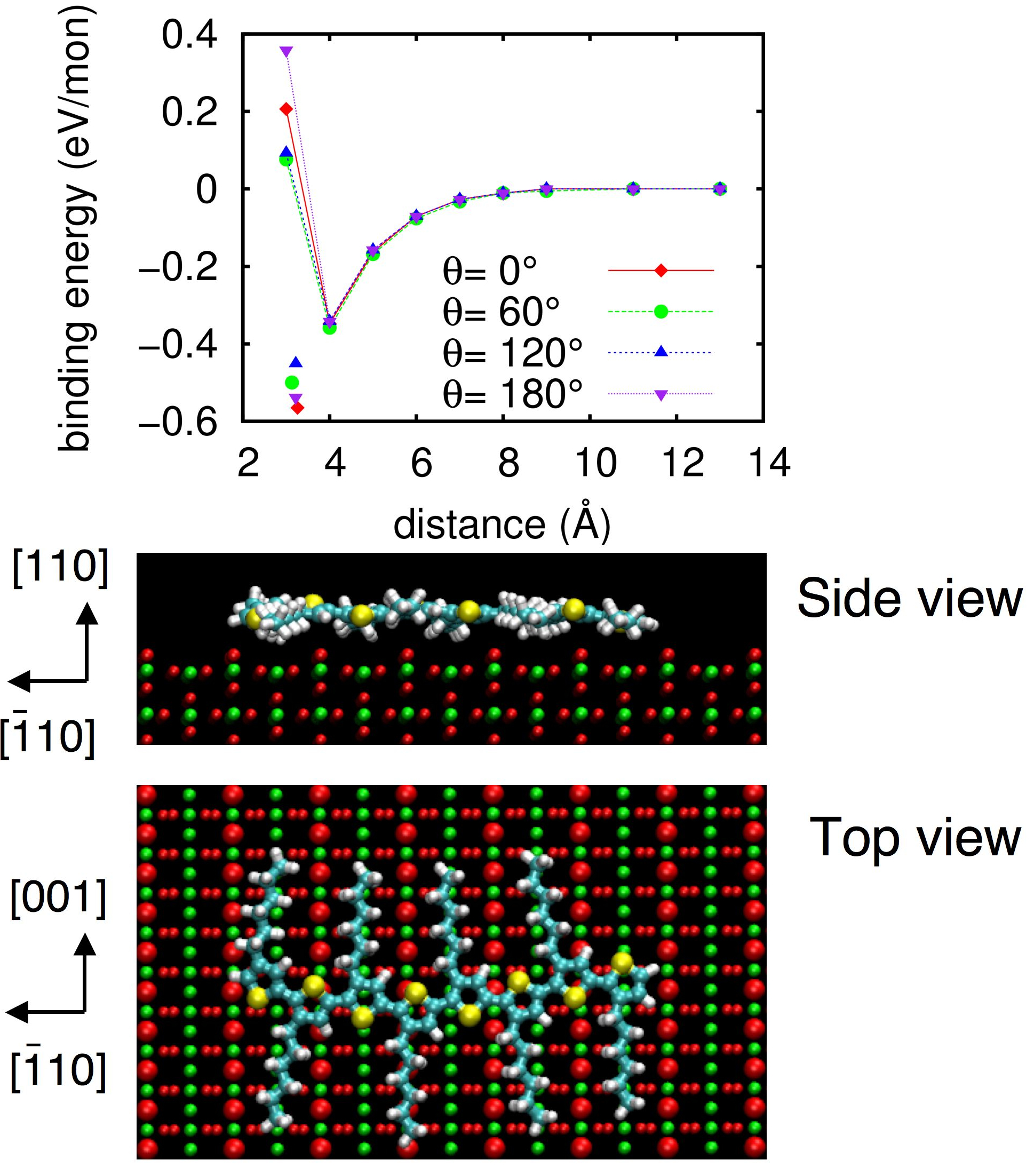}
\caption{\label{110_curve}
Top: Unrelaxed binding energy per monomer $U$ as  a function of the polymer-surface distance $d$ and the polymer 
orientation in the (110) plane. Top (middle) and  side view (bottom) of the polymer-surface bound state. Carbon, Hydrogen, Sulphur, Oxygen and Titanium atoms are represented in 
cyan, white, yellow, red and green, respectively. }
\end{figure}

In order to study the interaction between P3HT and titania surface we replaced the thiophene molecule  in the above analysis by 
 an oligothiophene (N-3hexylthiophene, N-3HT). We performed our analysis by considering  the case $N=8$ for which 
 the energy per monomer is close to the limit of an infinite chain.
The system is reported in \ref{110_curve}. The corresponding number of atoms was in this case prohibitively large for 
DFT calculations and the analysis was performed by only using MP calculations.

A relaxed oligothiophene was placed parallel to the surface at a given distance $d$ (the range $0.2-1.3$ nm).  For each $d$, we
considered different molecule orientations in the (110) plane defined by the angle $\theta$ formed  by the
polymer backbone and the $[\bar{1}10]$ direction. Positive $\theta$ values correspond to anti-clock wise rotation of the molecule.
The energy of the system $E(d,\theta)$ is calculated   on a mesh of values $(d,\theta)$ in order to explore the interaction basin.
For each $(d,\theta)$ the atoms where kept fixed (unrelaxed energy $U$).
The unrelaxed binding energy per monomer $U$ is here defined as  
$U(d,\theta) =(E(d,\theta)-E(\infty))/N$, where $E(\infty)$ corresponds to the energy of the unbound molecule-surface pair ($d \to \infty$). 
 The calculated $U(d,\theta)$ is reported in top panel of \ref{110_curve}.
 It is found that the titania surface gives rise to an attraction basin for the molecule with a minimum at 
$d \approx 0.4$ nm corresponding to a moderate (unrelaxed) binding energy $ U \approx -0.36$ eV per monomer.
 A slight dependence on the orientation $\theta$ is observed that increases at smaller distances.  
 The above analysis demonstrates the occurrence of an attractive interaction between the 
 P3HT and the surface acting up to distances  $d \approx 0.8$ nm.
 
 In order to identify  and to characterize the bound state we performed an extensive search of the absolute minimum.
 First, each molecule-surface system (for the whole set of ($d$,$\theta$) points of the mesh)  was locally optimized   (by using  a conjugate-gradient algorithm);
 we then identified the minimum  for each molecule-surface orientation $\theta$ and we annealed the corresponding configuration at $10$ K
for  $100$-ps. The resulting configuration was finally cooled down to T$=0$ K.
The corresponding energies are reported in \ref{110_curve} as points. Each color corresponds to different initial orientations.
Notably, all the identified bound states correspond to  molecule-surface distances
around $d_{min}=0.32$ nm and the largest binding energy ( $0.56$ eV/monomer) is found for the molecule aligned along the $[\bar{1}10]$ surface
crystallographic orientation.
The quantity $\eta_p = -U(d_{min}$) is the planar adhesion energy  and it is the work spent to detach the polymer from
the planar titania surface (per monomer of P3HT).

According to our analysis, the  energy decrease ($\approx 0.2$ eV/monomer) in the bound state with respect the unrelaxed molecule,
is mainly due to the optimization of the electrostatic energy. 
In particular, we observe that: 
(i) the more flexible  hexyl chains tend to minimize the unfavorable electrostatic interactions with the  negatively charged oxygen atomic rows and to maximize the favorable  electrostatic interactions with the titanium atoms.
(ii) the thiophene rings forming the stiff backbone of the molecule tend to sit on top of positively charge titanium atoms of the surface.
As a result, large distortions are observed in the molecule structure and the 
overall electrostatic gain is limited by the strain energy of the polymer.
The incommensurability between  the surface and  polymer gives rise to an aperiodic strain field within the molecule. 

\begin{figure}
\includegraphics[scale=1.1]{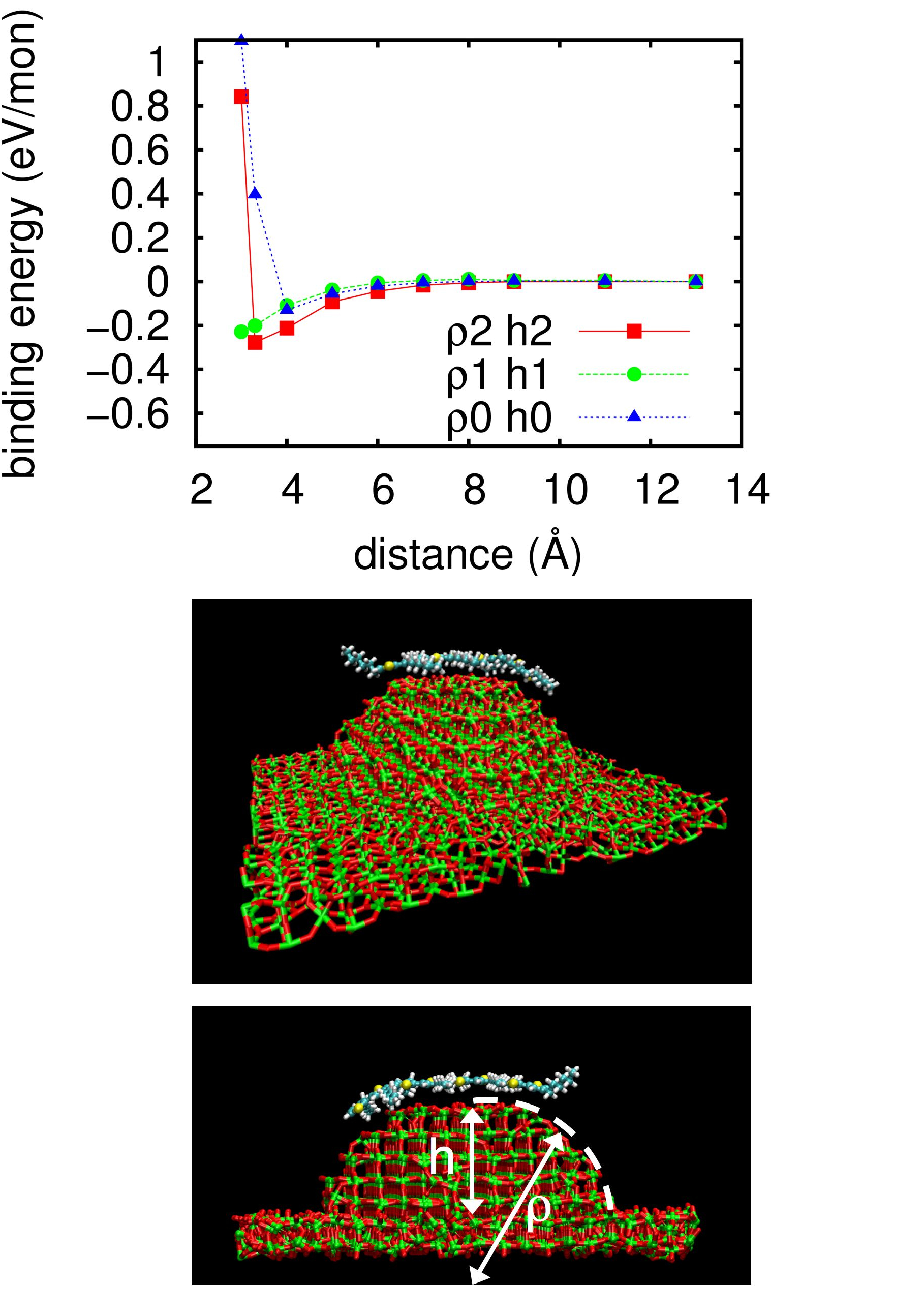}
\caption{  \label{HILL_curve}
Top: Unrelaxed binding energy per monomer, $U$,  as  a function of the polymer-surface distance $d$ and the polymer 
orientation in the (110) plane. Middle: Stick and balls representation of a  P3HT/TiO$_2$ system consisting of   
an oligothiophene placed on top of a spherical titania cap. Bottom: Side view of the P3HT/TiO$_2$ system, the white arrows represent the curvature radius 
$\rho$ and the height $h$ of the titania cap. Carbon, Hydrogen, Sulphur, Oxygen and Titanium atoms are represented in 
cyan, white, yellow, red and green, respectively. }
\end{figure}

The above results for the case of a planar surface may change when the surface is nanostructured,
as indeed commonly used in 
hybrid devices in order to improve the polymer adhesion.
The surface morphology of the titania film, the local charge of the surface, or the presence 
defects in the lattice can in principle 
strongly affect the polymer/TiO$_2$ interaction.
In order to elucidate these effects, we repeat the above analysis for the case of a model nanostructured titania,
  consisting in  a
 spherical caps placed on top of a planar  substrate. The cap is characterized by its  radius ($\rho$) and  height ($h$) 
 that mimic the local radius of curvature and roughness, respectively of a real nanostructure .
 An example  is reported in \ref{HILL_curve} for the case $\rho=0.23$ nm and $h=0.15$ nm.
 We considered nine nanostructures with radius curvature
 $\rho$ equal to $0.19$ nm, $0.23$ nm, $0.33$ nm and $h$ equal to $0.9$ nm, $0.15$ nm, $0.24$ nm.
 Hereafter the lengths scale  will be normalized to the oligothiophene length ($3.2$ nm). The case $\rho=1$ corresponds to
 a curvature radius equal to the molecule length.
For a given curvature radius $\rho$, the height $h$ must be such that $0 \le \rho \le 2\rho$.
 For $\rho \ge 2\rho$ a spherical cluster is obtained. 
 The atomistic models were generated according to the following procedure:
(i) each nanostructure of given $\rho$ and $h$ was carved out from a monocrystal rutile by strictly preserving the charge neutrality of the system;
(ii) each system was annealed at $10$ K for $100$-ps and further relaxed by a conjugate-gradients algorithm.
During the minimization procedure, large atomic relaxations are observed both on  the surface of the  cap and on the planar substrate that
 depend on the actual curvature and height.  

\begin{figure}
\includegraphics[scale=1.0]{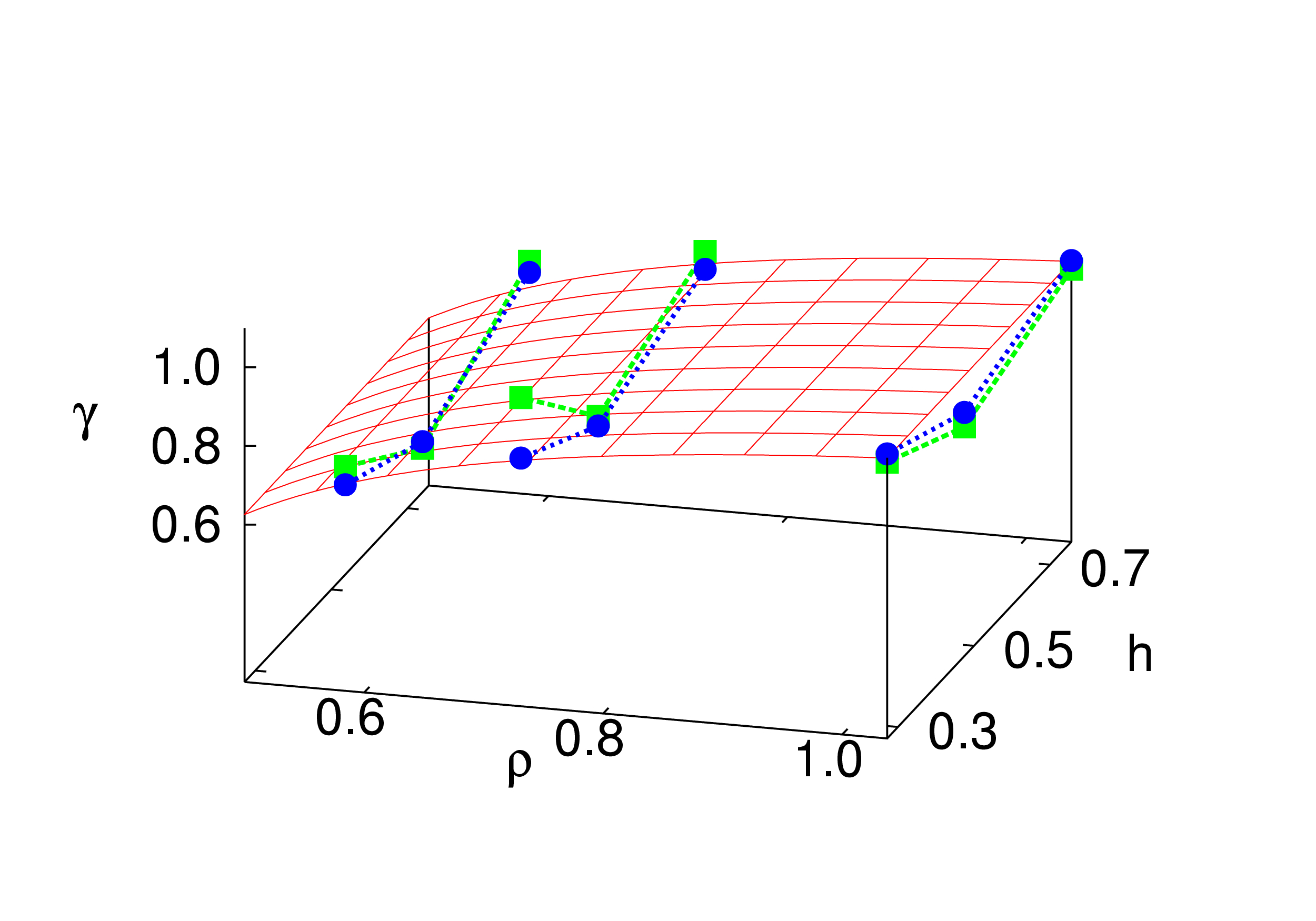}
\caption{  \label{binding_curve}
Adhesion energy $\gamma$  as a function of the local curvature radius and roughness of the titania surface. The atomistic data are represented as green squares; blue circles are correspond to the calculations based on the continuum model.
}
\end{figure}

The interaction between the oligomer and the fully relaxed nanostructures was studied by placing the molecule on top of the cap and
 by performing similar calculations as in the case of a planar interface.
 According to our calculations, all the nanostructures  give rise to an effective attraction basin for the molecule (\ref{HILL_curve}, top). The
corresponding  bound configuration 
and  binding energy were calculated as a function of ($\rho$-$h$).
 In \ref{binding_curve} we report   
$
\gamma= -U/\eta_p 
$
i.e. the adhesion energy (per monomer) normalized to the value corresponding  to  planar titania surface (green squares).

The first important fact is that the calculated adhesion energy $\gamma$ is affected by the morphology of the surface and a nonmonotonic dependence on 
$h$ and $\rho$ is observed.
By definition, as $\rho \to \infty$ the nanostructure tends to a planar interface and the adhesion must be close to one ( $\gamma \to 1$).
 This is in fact the case of the calculated data, where it is observed that on average $\gamma$ increases with $\rho$ and it approaches the
 planar value.
 
We attribute the $\rho$ dependence of $\gamma$ to an elastic effect due to the polymer bending.
The adhesion  on a curved surface requires a  bending proportional to the surface curvature (see \ref{binding_curve}) that corresponds to an elastic
energy cost that reduces $\gamma$.
As the radius $\rho$ decreases, the strain similarly increases and the adhesion is less effective.
It is known that the strain energy of a bent plate  increases quadratically with the curvature $ \sim \delta ^3/\rho^2$ where
$\delta$ is the width of the plate.\cite{lavoro_elastico}
Accordingly, we expect that
\begin{equation}
\gamma(\rho) = 1 - \alpha \rho ^{-2}
\end{equation}
 where $\alpha$ is proportional to the elastic torsional constant of the polymer.
 The above model (red surface, \ref{binding_curve}) is able to fit the overall behavior of the atomistic data by using a constant  $\alpha=0.09$.
This confirms that the $\rho$ dependence of $\gamma$ is mainly due to an  elastic effect.
 On the contrary, the elastic model cannot explain the calculated dependence of $\gamma$  upon the  height $h$ of the nanostructure.
 In fact, for each curvature $\rho$,  the  
 adhesion is a function of $h$ with  a minimum around $h=0.5$ (half the polymer length) while the model predict a constant
 (red surface).
 The $h$ dependence is therefore explained in terms of atomistic effect due to the local charge of the cap.
 This is proved by calculating the local charge of cap for all the nanostructures.
 Though the whole system is electrically neutral, a net charge can be found in the cap, depending on the local titania stoichiometry.
In particular, for a given $\rho$, the cap stoichiometry depends on actual  number of atomic layers that, in turn depends on the height  $h$  of the cap.


This effect can be included in the above continuum model $\gamma(\rho)$ by adding an extra term 
related to the electrostatic interaction between the molecule and the local charge of the nanostructure.
In particular, we expect that the local charge $Q(h,\rho)$ of the cap interacts with the
dipole moment of the molecule giving rise to a dipole-monopole term: $ Q(h,\rho)  \mu_p  R_{eff}^{-2}  $.
 $R_{eff}$ is the effective distance between the molecule and the center of the cap and $\mu_p$ is the dipole projection on the 
 plane perpedicular to the monopole-dipole direction. 

By assuming that  $\mu_p$ and $R_{eff}$  does not  strongly depend on the actual surface morphology the above  electrostatic term  is simply proportional to  $Q$ and we can write:
\begin{equation}
\gamma(\rho,h) = 1 - \alpha \rho^{-2} - \beta Q(\rho,h) 
\end{equation}
The constant $\beta = (N \eta_p  )^{-1} R_{eff}^{-2}  \mu_p$ is adjusted
 to reproduce the atomistic data for $\gamma$.
 The actual $\mu_p$  can be calculated from the atomic coordinates and partial charges of the molecules
 in the bound states ($\mu_p \sim 1.4$ a.u.) 
 and the effective distance is of the order of 
 $R_{eff} \sim 1.5 $ nm that is consistent with the size of the caps. 
 The positive value of $\beta$ stands for a repulsive contribution.

The above model is able to  reproduce the dependence of the adhesion energy $\gamma$ on both  $\rho$ and $h$.
In  \ref{binding_curve} the data calculated by the above 
continuum model are reported as blue symbols.
An overall aggreement is found and in particular the dependence on $h$ is captured.
The deviations of the calculated data from the model are due to other atomistic features (not taken into account in the model), such as the VdW 
dispersion forces, the local distortion of the polymer, atomic scale details that cannot be described by the simple model
here proposed.
Nevertheless,  according to our findings, these contributions are minor corrections (few percents) to the $\gamma(\rho,h)$ model. 
 This confirms the interpretation of the adhesion mainly  in terms of electrostatic plus elastic effects. 
\subsection{Conclusions}
In conclusion, we have proved that the adhesion of P3HT on titania is dominated by electrostatic contributions. 
In addition, our calculations suggest that the adhesion in the case of
a nanostructured titania film differs sizeably by the case of a planar surface.
Nevertheless, we found that the nanomorphology does not necessarily increase the polymer adhesion with respect to the planar case.
In particular, when the curvature radius is much smaller than the average polymer chains length, 
 the molecular strain turns out to be detrimental for the adhesion efficiency. 
A better adhesion is predicted when  curvature $\rho$ is greater than the average polymer chain length.
Finally, we proved that it is important to take into account the local charge of the titania nanostructure in order to predict
the actual adhesion energy.

\begin{acknowledgement}
We acknowledge computational support by COSMOLAB (Cagliari, Italy) and CASPUR (Rome, Italy). 

\end{acknowledgement}

\begin{suppinfo}
 Additional materials (movies and images) is
availble free of charge via the Internet at
\url{http://slacs.dsf.unica.it/index.php?option=com_content&view=article&id=74&Itemid=94}.''
\end{suppinfo}

\bibliography{achemso}


\end{document}